\documentclass[a4paper]{article}
\usepackage[left=1in,top=1in,right=1in,bottom=1in]{geometry}
\usepackage[cmex10]{amsmath}
\usepackage{amssymb}
\usepackage{setspace}
\newcommand{\R}{{\mathbf R}}
\newcommand{\C}{{\mathbf C}}
\newcommand{\T}{{\mathbf T}}

\newcommand{\calB}{\mathcal{B}}
\newcommand{\calC}{\mathcal{C}}
\newcommand{\calD}{\mathcal{D}}
\newcommand{\calF}{\mathcal{F}}
\newcommand{\calG}{\mathcal{G}}
\newcommand{\calH}{\mathcal{H}}
\newcommand{\calK}{\mathcal{K}}
\newcommand{\calL}{\mathcal{L}}
\newcommand{\calM}{\mathcal{M}}
\newcommand{\calO}{\mathcal{O}}
\newcommand{\calS}{\mathcal{S}}
\newcommand{\calU}{\mathcal{U}}

\renewcommand{\Pr}[1]{\mathrm{Pr}\left(#1\right)}
\newcommand{\E}[1]{{\mathbf E}\left(#1\right)}
\newcommand{\I}[1]{\mathbf{1}\left\{#1\right\}}
\newcommand{\dif}{\textrm{d}}
\newcommand{\Av}[2]{\mathcal{A}_{#1}\left(#2\right)}
\newcommand{\AV}{\mathcal{A}}

\newcommand{\by}{\times}
\newcommand{\Mat}[2]{\calM_{#1}^{#2}}
\newcommand{\Diag}[2]{\calD_{#1}^{#2}}
\newcommand{\CvM}[2]{\calK_{#1}^{#2}}
\newcommand{\Uni}[1]{\calU_{#1}}
\newcommand{\diag}[1]{\textnormal{diag}\left(#1\right)}
\newcommand{\Tr}[1]{\textrm{Tr}\left(#1\right)}

\title{On the Symmetries and the Capacity Achieving Input Covariance Matrices of Multiantenna Channels\footnote{This paper is an extended version of the paper with same title presented at the 2016 IEEE International Symposium on Information Theory.}}
\author{Mario Diaz\thanks{M. Diaz is with the Department of Mathematics and Statistics, Queen's University, Kingston,
ON Canada. e-mail: 13madt@queensu.ca}}
\date{}

\begin{document}

\maketitle

\begin{abstract}
In this paper we study the capacity achieving input covariance matrices of a single user multiantenna channel based solely on the group of symmetries of its matrix of propagation coefficients. Our main result, which unifies and improves the techniques used in a variety of classical capacity theorems, uses the Haar (uniform) measure on the group of symmetries to establish the existence of a capacity achieving input covariance matrix in a very particular subset of the covariance matrices. This result allows us to provide simple proofs for old and new capacity theorems. Among other results, we show that for channels with two or more standard symmetries, the isotropic input is optimal. Overall, this paper provides a precise explanation of why the capacity achieving input covariance matrices of a channel depend more on the symmetries of the matrix of propagation coefficients than any other distributional assumption.
\end{abstract}

\section{Introduction and Preliminaries}
\label{Section:Introduction}

In the past, capacity theorems for various single user multiantenna channels have been found, e.g. \cite{Te99,VeSiVa03,LoTuVe06,DiPe15}. These capacity theorems rely on different assumptions for the matrix of propagation coefficients of the channel: independent identically distributed Gaussian entries, independent columns each being symmetric around zero, unitary rotations of the previous ones, etc. However, as we shall see, the symmetries of the matrix of propagation coefficients of the channel are the heart of the matter. In our context, these symmetries are unitary matrices that, under conjugation, leave invariant the distribution of the product of the matrix of propagation coefficients and its conjugate transpose, see equation (\ref{eq:DefGroupSymmetries}). An analysis based on these symmetries does not depend on moment conditions, correlation assumptions, or distributional requirements for the propagation coefficients. As a consequence, many common models in the literature can be analyzed within a single framework. In this paper we study the capacity achieving input covariance matrices of single user multiantenna channels based on the aforementioned symmetries.

We consider a single user multiantenna channel where the receiver and the transmitter use $M$ and $N$ antennas respectively. Moreover, we assume that channel state information at the receiver and channel distribution information at the transmitter are available. Under the assumption of a linear vector memoryless channel ${\bf y}=H{\bf x}+{\bf n}$ \cite[Sec. 1.1]{TuVe2004}, the behavior of the channel is encoded in its matrix of propagation coefficients, an $M\by N$ random matrix $H$. For notational simplicity we will assume that both the transmitter power and the signal-to-noise ratio equal one. Observe that this can be done by incorporating the quotient of the square roots of the transmitter and noise powers to the matrix of propagation coefficients of the channel. For any matrix $A$, we denote by $A_{i,j}$ its $i,j$-entry and by $\Tr{A}=\sum_i A_{i,i}$ its trace. Let $\CvM{N,1}{\C}$ denotes the set of $N\by N$ covariance matrices with complex entries and unital trace. The mutual information $I_H(Q):=I({\bf x}^Q;({\bf y},H))$ attained by a circularly symmetric complex Gaussian vector ${\bf x}^Q$ with covariance $Q\in\CvM{N,1}{\C}$ is given by 
\begin{equation}
\label{eq:DefMutualInformation}
I_H(Q) = \E{\log\det\left({\rm I}_M + HQH^*\right)},
\end{equation}
and the ergodic capacity $C_H$ of $H$ by
\begin{equation}
\label{eq:DefCapacity}
C_H = \sup_{Q\in\CvM{N,1}{\C}} I_H(Q).
\end{equation}
Here ${\rm I}_M$ denotes the $M\by M$ identity matrix and ${\mathbf E}$ denotes the expectation with respect to the distribution of $H$. A matrix $Q\in\CvM{N,1}{\C}$ such that $I_H(Q)=C_H$ is called a capacity achieving input covariance matrix (CAICM) for $H$.

An $N\by N$ matrix $V$ is called unitary if $VV^\ast=V^\ast V={\rm I}_N$ where $V^\ast$ is the conjugate transpose of $V$. We denote the multiplicative group of $N\by N$ unitary matrices by $\Uni{N}$ and the set of $M\by N$ (resp. $N\by N$) complex matrices by $\Mat{M\by N}{\C}$ (resp. $\Mat{N}{\C}$). Let $\calF$ be a finite multiset\footnote{A set-like object in which repeated elements are allowed.} with elements in $\Uni{N}$. Motivated by \cite[Lemma 1]{DiPe15}, we consider the average operator $\AV_\calF:\Mat{N}{\C}\to\Mat{N}{\C}$ with respect to $\calF$,
\begin{equation}
\label{eq:DefAvF}
\Av{\calF}{A} = \frac{1}{|\calF|} \sum_{F\in\calF} FAF^*.
\end{equation}
Recall that the set of covariance (or positive semidefinite) matrices is a closed cone. In particular, $\AV_\calF$ sends $\CvM{N,1}{\C}$ into itself. We use $X\stackrel{\calL}{=}Y$ to denote equality in distribution, i.e., $\Pr{X\in\calB}=\Pr{Y\in\calB}$ for every Borel set $\calB\subset\Mat{M\by N}{\C}$. The following proposition, which is a reformulation of Telatar's observation based on Jensen's inequality and the concavity of the log-det over the cone of positive semidefinite matrices \cite{Te99}, states that certain average operators do not decrease the mutual information.\\

\noindent{\bf Proposition 1} Let $H$ be an $M\by N$ random matrix. Suppose that $\calF\subset\Uni{N}$ is a finite multiset such that $F^*H^*HF\stackrel{\calL}{=}H^*H$ for all $F\in\calF$. Then $I_H(Q) \leq I_H(\Av{\calF}{Q})$ for all $Q\in\CvM{N,1}{\C}$. In particular, if $Q_0$ is a CAICM for $H$ then $\Av{\calF}{Q_0}$ is also a CAICM for $H$.\\

Proposition 1 is a key ingredient in a whole family of capacity theorems for multiantenna channels \cite{Te99,LoTuVe06,DiPe15}. In general\footnote{In Telatar's work \cite[Th. 1]{Te99}, a sightly different study of the covariance matrices was performed. However, as a consequence of \cite[Th. 1]{DiPe15} and its proof, an analysis using the following strategy is also possible.}, the strategy is to find a finite family of finite sets $\calF_1,\ldots,\calF_n$ such that they satisfy the hypothesis in Proposition 1 and they make the set $\calC=\AV_{\calF_n}\circ\cdots\circ\AV_{\calF_1}(\CvM{N,1}{\C})$ as small as possible. Then, by the previous proposition, there exists a CAICM for $H$ in $\calC$. Here $f\circ g$ means the composition of the functions $g$ and $f$. Notice that $\AV_{\calF_n}\circ\cdots\circ\AV_{\calF_1}$ can be replaced by $\AV_\calF$ for an appropriate multiset $\calF$. Up to date, the procedure for finding appropriate sets $\calF_1,\ldots,\calF_n$ seems to be tricky or ingenious and, when the resulting set $\calC$ is not the set containing only the normalized identity $\frac{1}{N}{\rm I}_N$, there is no guarantee that $\calC$ cannot be further reduced using some extra sets $\calF_{n+1},\ldots,\calF_{n+m}$.

Relying on the concept of the group of symmetries of the matrix of propagation coefficients (see equation (\ref{eq:DefGroupSymmetries})), the main result of this paper provides a set where a CAICM for $H$ exists. This result does not depend on any ingenious guessing and provides a set smaller than or equal to any set obtained using a finite multiset, i.e., this non-guessing result is at least as good as any method based on the strategy in the previous paragraph. The non-guessing nature of this result leads to simple proofs to old and new capacity theorems. One of these new theorems shows why even in a channel with very few symmetries, the isotropic input is optimal (see Proposition 3).

This paper has four sections apart from this. In Section \ref{Section:MainResults} we present the main results and contributions of this paper. We apply them to either improve the statements or simplify the proofs of known capacity theorems in Section \ref{Section:FurtherApplications}. In Section \ref{Section:OptimalityMainTheorem} we prove the optimality of our main theorem: we show that in general, based on the symmetries of the matrix of propagation coefficients of the channel, it is impossible to do better than Theorem 1. In Section \ref{Section:Conclusion} we make some concluding remarks.

\section{Main results}
\label{Section:MainResults}

Let $H$ be the $M\by N$ random matrix associated to a single user multiantenna channel as the ones described in the previous section. We define the {\it group of symmetries} $\calG(H)$ of $H$ by
\begin{equation}
\label{eq:DefGroupSymmetries}
\calG(H) := \left\{ V\in\Uni{N} \mid V^*(H^*H)V\stackrel{\mathcal{L}}{=} H^*H \right\}.
\end{equation}
Despite the name, $\calG(H)$ depends on $H$ only through the distribution of $H^*H$. If $HV \stackrel{\mathcal{L}}{=} H$ for some $V\in\Uni{N}$, it is straightforward to see that $V\in\calG(H)$. This criterion will be used repeatedly in Section \ref{Section:FurtherApplications}.

The following technical, yet intuitive, lemma plays a key role in this paper. It allows to bring in the theory of the Haar measure, making possible to define the average operator with respect to any closed subgroup of $\Uni{N}$, see equation (\ref{eq:DefAvExt}).\\

\noindent{\bf Lemma 1} Let $H$ be an $M\by N$ random matrix. Then its group of symmetries $\calG(H)$ is a non-empty closed subgroup of $\Uni{N}$.\\

\noindent{\it Proof.} See Appendix \ref{Appendix:Lemma1}. $\square$\\

Suppose that $\calF$ is a closed subgroup of the compact group $\Uni{N}$. By the theory of the Haar measure on compact topological groups \cite[Ch. VI]{Lo53}, there exists a unique probability measure $\mu_\calF$ on the Borel sets of $\calF$ such that $\mu_\calF(\calB)=\mu_\calF(F\calB)=\mu_\calF(\calB F)$ for every Borel set $\calB\subset\calF$ and every $F\in\calF$. The measure $\mu_\calF$ is the so-called Haar measure on $\calF$. Intuitively speaking, the Haar measure is nothing but the uniform distribution on the corresponding space, e.g., the Haar measure on the circle is the (properly normalized) measure determined by arc-length. We define the average operator $\AV_{\calF}:\Mat{N}{\C}\to\Mat{N}{\C}$ with respect to $\calF$ by
\begin{equation}
\label{eq:DefAvExt}
\Av{\calF}{A} := \int_{\calF} FAF^* \dif\mu_{\calF}(F).
\end{equation}
When $\calF$ is a finite subgroup of $\Uni{N}$, the previous definition of $\AV_\calF$ coincides with the one in equation (\ref{eq:DefAvF}). In fact, the extended average operator does the same kind of operation as the average operator, but using a continuous set instead of a finite one. Observe that $\AV_\calF$ sends $\CvM{N,1}{\C}$ into itself. Let $U$ be a random matrix distributed according to the Haar measure on $\calF$. The matrix integral in equation (\ref{eq:DefAvExt}) can be computed from the joint moments of order two of the entries of $U$. Specifically, for all $1\leq i,j\leq N$,
\begin{equation}
\label{eq:AvMoments}
\Av{\calF}{A}_{i,j} = \E{UAU^*}_{i,j} = \sum_{k,l=1}^N A_{k,l} \E{U_{i,k}\overline{U_{j,l}}}.
\end{equation}
Usually, we can compute the moments in the previous equation directly from the multiplication invariance property of the Haar measure, without computing any integral (see Section \ref{Section:FurtherApplications}). This observation has important consequences in practice: it replaces the rather abstract definition of the average operator in equation (\ref{eq:DefAvExt}) with the simpler expression in equation (\ref{eq:AvMoments}).

By Lemma 1, $\calG(H)$ is a closed subgroup of $\Uni{N}$ and thus the average operator with respect to $\calG(H)$ is well defined. Our main result is as follows.\\

\noindent{\bf Theorem 1} Let $H$ be an $M\by N$ random matrix and $\calG(H)$ be its group of symmetries. Then,\\
\indent a) There exists a CAICM for $H$ in $\Av{\calG(H)}{\CvM{N,1}{\C}}$;\\
\indent b) $\Av{\calG(H)}{\CvM{N,1}{\C}} \subset \Av{\calF}{\CvM{N,1}{\C}}$ whenever $\calF$ is a finite multiset with elements in $\calG(H)$ or a closed subgroup of $\calG(H)$.\\

\noindent{\it Proof.} See Appendix \ref{Appendix:Thm1}. $\square$\\

Part a) establishes the existence of a CAICM for $H$ in the set $\{\Av{\calG(H)}{Q} \mid Q\in\CvM{N,1}{\C}\}$. Since $\calG(H)$ is the biggest set whose elements satisfy the hypothesis in Proposition 1, part a) can be regarded as the most general version of Proposition 1. Recall that the objective of a capacity theorem is to provide the smallest set where a CAICM for $H$ is guaranteed to exit. Part b) says that the average operator with respect to the group of symmetries produces a set smaller than or equal to any set produced by a finite multiset $\calF$. Overall, Theorem 1 avoids any ingenious guessing and produces a result at least as good as the best ingenious guessing. Furthermore, in Section \ref{Section:OptimalityMainTheorem} we shall show that there are examples where it is impossible to do better than Theorem 1 based solely on the symmetries of the matrix of propagation coefficients of the channel. This proves the optimality of our main result within the techniques based on the aforementioned symmetries.

A straightforward consequence of Theorem 1 part b), which cannot be proved from Proposition 1 or  even Theorem 1 part a), is Proposition 2 below. This proposition plays an important role in our main application, Proposition 3.\\

\noindent{\bf Proposition 2} Let $H$ be an $M\by N$ random matrix. If $\calF_k$ is either a finite multiset with elements in $\calG(H)$ or a closed subgroup of $\calG(H)$ for $1\leq k\leq K$, then there exists a CAICM in the set $\bigcap_{k=1}^K \Av{\calF_k}{\CvM{N,1}{\C}}$.\\

\noindent{\it Proof.} See Appendix \ref{Appendix:Prop2}. $\square$\\

Let $\Diag{N}{\T}$ be the set of $N\by N$ diagonal matrices with values on $\T=\{z\in\C \mid |z|=1\}$. By standard results in linear algebra, any $V\in\Uni{N}$ can be written as $V=WDW^*$ for some $W\in\Uni{N}$ and $D\in\Diag{N}{\T}$.  We say that $V$ is a {\it standard symmetry} if $\left\{1,\frac{\arg(D_{1,1})}{2\pi},\ldots,\frac{\arg(D_{N,N})}{2\pi}\right\}$ are rationally independent, i.e., if
\begin{equation*}
q_0 + q_1 \frac{\arg(D_{1,1})}{2\pi} + \cdots + q_N \frac{\arg(D_{N,N})}{2\pi}=0
\end{equation*}
for some rational numbers $q_0,\ldots,q_N$, then $q_0=q_1=\cdots=q_N=0$. Even though the previous condition seems rather technical, it can be shown, using the Weyl integration formula \cite[Th. IX.9.1]{Si96}, that the set of standard symmetries is a set of measure one with respect to the Haar measure on $\Uni{N}$. Thus, a standard symmetry might be thought as a typical element in $\Uni{N}$. In this context, our main application is the following proposition which gives some sufficient conditions for the optimality of the isotropic input. Since these conditions rely only on the symmetries of the matrix of propagation coefficients, they are essentially different from those given in \cite[Prop. 1]{LoTuVe06} for example. Another conditions for the optimality of the isotropic input, that depend only on the symmetries of the matrix of propagation coefficients, are given in Corollary 5 below.\\

\noindent{\bf Proposition 3} Suppose that $H$ is an $M\by N$ random matrix. If the group of symmetries $\calG(H)$ of $H$ has two standard symmetries $V_1=W_1D_1W_1^*$ and $V_2=W_2D_2W_2^*$ such that $W_{i,j}\neq0$ for all $1\leq i,j\leq N$ where $W=W_1^*W_2$, then the normalized identity is a CAICM for $H$.\\

\noindent{\it Proof.} See Appendix \ref{Appendix:Prop3}. $\square$\\

In other words, the isotropic input is optimal for any channel with at least two essentially different standard symmetries. It is possible to prove, along the same lines as in \cite[Sec. 6]{CoMa14}, that two independent random matrices distributed according to the Haar measure on the unitary matrices satisfy the hypothesis of the previous proposition a.s. Thus, if a channel has a typical pair of unitary matrices as symmetries, in the sense of coming as a realization of a pair of independent Haar unitary random matrices, then Proposition 3 implies that the isotropic input is optimal.

As a byproduct of our investigations, we obtained the following proposition. This proposition is used to prove Theorem 1, but we consider that it might be of general interest by itself. Recall the expressions for the mutual information and the capacity in equations (\ref{eq:DefMutualInformation}) and (\ref{eq:DefCapacity}) respectively. In general, it is possible to have $C_H=\infty$. The following proposition establishes the equivalence between the finiteness of the capacity and the continuity of $I_H(\cdot)$. Furthermore, it shows that the physically uninteresting case $C_H=\infty$ is also theoretically uninteresting.\\

\noindent{\bf Proposition 4} Suppose that $H$ is an $M\by N$ random matrix. Then the following are equivalent:\\
\indent a) $\E{\log(1+\|H\|)}$ is finite;\\
\indent b) $I_H(\cdot)$ is continuous on $\CvM{N,1}{\C}$;\\
\indent c) $C_H$ is finite.\\
In addition, $C_H=\infty$ if and only if $I_H(\frac{1}{N}{\rm I}_N)=\infty$.\\

\noindent{\it Proof.} See Appendix \ref{Appendix:Prop4}. $\square$\\

Here $\|A\|$ denotes the Frobenius norm $\|A\| = \sqrt{\Tr{A^*A}}$. The Frobenius norm in part a) can be replaced by any other norm. Indeed, this is because of the inequality $\log(1+\alpha x)\leq \log(1+x)+\log(1+\alpha)$ for $\alpha,x\geq0$ and the equivalence of norms in finite dimensional vector spaces \cite[Corollary 3.14]{La93}. Also, an application of Jensen's inequality shows that if $\sum_{i,j} \E{|H_{i,j}|^2} < \infty$, then condition a) is satisfied. Thus, any channel with propagation coefficients having finite variance has finite capacity.

\section{Further Applications}
\label{Section:FurtherApplications}

In this section we apply our main results to improve the statements or simplify the proofs of capacity theorems already known in the literature. Particular attention should be put to Corollary 6, a stronger version of the main theorem in \cite{VeSiVa03}. The following applications are divided according to their underlying symmetries: unitary, diagonal and block symmetries.

Through this section, for a square matrix $M$, we let $\Delta(M)$ be the diagonal matrix that has the same diagonal elements as $M$. Also, we let $\diag{d_1,\ldots,d_N}$ be the diagonal matrix with $d_1,\ldots,d_N$ as its diagonal elements. By a Gaussian random matrix we mean a random matrix whose entries are independent identically distributed circularly symmetric complex Gaussian random variables. Only one property of the Gaussian random matrices is used below: if $H$ is a Gaussian random matrix, then $VHW^*\stackrel{\calL}{=}H$ for all $V\in\Uni{M}$ and $W\in\Uni{N}$ \cite[Lemma 5]{Te99}. Thus, our approach generalizes straightforwardly to any unitarily invariant random matrix model.

\subsection{Unitary symmetries}
\label{Subsection:UnitarySymmetries}

In this section we study the capacity achieving input covariance matrices of channels whose matrices of propagation coefficients have the largest possible group of symmetries: the whole unitary group $\Uni{N}$.

Let $U$ be an $N\by N$ random matrix distributed according to the Haar measure on $\Uni{N}$. The moments of order two of the entries of $U$ are given by
\begin{equation}
\label{eq:MomentsHaarUnitary}
\E{U_{i,k}U_{j,l}}=0\textnormal{ and }\E{U_{i,k}\overline{U_{j,l}}} = \frac{1}{N}\delta_{i,j}\delta_{k,l}
\end{equation}
for all $1\leq i,j,k,l\leq N$ where $\delta_{p,q}$ equals one if $p=q$ and zero otherwise. The previous equation can be derived from the multiplication invariance property defining the Haar measure \cite[Lemma 1.1 and Prop. 1.2]{HiPe00}.\\

\noindent{\bf Lemma 2} For every $A\in\Mat{N}{\C}$, $\Av{\Uni{N}}{A}=\frac{\Tr{A}}{N}{\rm I}_N$. In particular,
\begin{equation*}
\Av{\Uni{N}}{\CvM{N,1}{\C}} = \left\{\frac{1}{N}{\rm I}_N\right\}.
\end{equation*}

\noindent{\it Proof.} See Appendix \ref{Appendix:Lemma2}. $\square$\\

Telatar's theorem \cite[Th. 1]{Te99} is then a corollary to the previous lemma and Theorem 1 part a).\\

\noindent{\bf Corollary 1} If $H$ is an $M\by N$ Gaussian random matrix, then $\frac{1}{N}{\rm I}_N$ is a CAICM for $H$.\\

\noindent{\it Proof.} By \cite[Lemma 5]{Te99} we know that $H \stackrel{\calL}{=} HV$ for every $V\in\Uni{N}$. In particular, $\calG(H) = \Uni{N}$. By Lemma 2 and part a) of Theorem 1, there is a CAICM for $H$ in
\begin{equation*}
\Av{\calG(H)}{\CvM{N,1}{\C}}=\left\{\frac{1}{N}{\rm I}_N\right\}.\;\square
\end{equation*}

\subsection{Diagonal symmetries}
\label{Subsection:DiagonalSymmetries}

After the unitary group, the most natural group of symmetries to study is the diagonal unitary matrices $\Diag{N}{\T}\subset\Uni{N}$. In this section we study this group of symmetries and its rotations.

Consider a unitary matrix $W$ in $\Uni{N}$. Observe that $W\Diag{N}{\T}W^*$ is a closed subgroup of $\Uni{N}$. Let $U=\diag{u_1,\ldots,u_N}$ be an $N\by N$ random matrix such that $u_1,\ldots,u_N$ are independent random variables uniformly distributed on $\T$. It is a routine exercise to show that $(WUW^*)(WDW^*)\stackrel{\calL}{=} WUW^*$ for every $D\in\Diag{N}{\T}$. Since the Haar measure is the unique multiplication invariant probability measure, we conclude that $WUW^*$ is distributed according to the Haar measure on $W\Diag{N}{\T}W^*$. Let $\Diag{N,1}{\R_+}$ be the set of $N\by N$ diagonal matrices with non negative entries and unital trace.\\

\noindent{\bf Lemma 3} Let $W$ be a unitary matrix. For every $A\in\Mat{N}{\C}$, $\Av{W\Diag{N}{\T}W^*}{A}= W\Delta(W^*AW)W^*$. In particular,
\begin{equation*}
\Av{W\Diag{N}{\T}W^*}{\CvM{N,1}{\C}} = W\Diag{N,1}{\R_+}W^*.
\end{equation*}

\noindent{\it Proof.} See Appendix \ref{Appendix:Lemma3}. $\square$\\

As a consequence of the previous lemma, we obtain the following corollaries.\\

\noindent{\bf Corollary 2} \cite[Th. 1]{LoTuVe06} Let $H=W_M \tilde{H} W_N$ be an $M\by N$ random matrix such that $W_M$ and $W_N$ are deterministic unitary matrices and $\tilde{H}$ is an $M\by N$ random matrix with independent columns the distribution of whose entries is jointly symmetric with respect to zero. There is a CAICM for $H$ of the form $W_N^* D W_N$ for some diagonal matrix $D\in\Diag{N,1}{\R_+}$.\\

\noindent{\it Proof.} By assumption, the distribution of $\tilde{H}$ is invariant under right multiplication by diagonal unitary matrices. In particular, $HW_N^*DW_N \stackrel{\mathcal{L}}{=} H$ and thus $W_N^*\Diag{N}{\T}W_N\subset\calG(H)$. From Lemma 3 and Theorem 1, there is a CAICM for $H$ in
\begin{equation*}
\Av{\calG(H)}{\CvM{N,1}{\C}}\subset\Av{W_N^*\Diag{N}{\T}W_N}{\CvM{N,1}{\C}}=W_N^*\Diag{N,1}{\R_+}W_N.\;\square
\end{equation*}

\noindent{\bf Corollary 3} \cite[Th. 3]{LoTuVe06} Let $H=c_Mc_N^*$ where $c_M$ and $c_N$ are independent column vectors each having independent random entries whose distributions are symmetric with respect to zero. A CAICM for $H$ is diagonal.\\

\noindent{\it Proof.} Since the entries of $c_N$ are independent and symmetric with respect to zero, we have that $c_N^*D\stackrel{\calL}{=}c_N^*$ for every $D\in\Diag{N}{\T}$. By the independence between $c_M$ and $c_N$, we obtain that $\Diag{N}{\T}\subset\calG(H)$. By Lemma 3 and Theorem 1, there is a CAICM for $H$ in
\begin{equation*}
\Av{\calG(H)}{\CvM{N,1}{\C}}\subset\Av{\Diag{N}{\T}}{\CvM{N,1}{\C}}=\Diag{N,1}{\R_+}.\;\square
\end{equation*}

\subsection{Signed permutation symmetries}
\label{Subsection:SignedPermutationSymmetries}

In this section we recall some basic facts about the Haar measure on finite groups. In addition, we examine a remarkable finite group: the signed permutation matrices.

Let $\calG$ be a finite subgroup of $\Uni{N}$. If $U$ is a random variable uniformly distributed in the (finite) set $\calG$, then
\begin{align*}
\Pr{UG \in \calF} &= \Pr{U\in \calF G^{-1}} = \frac{|\calF G^{-1}|}{|\calG|} = \frac{|\calF|}{|\calG|} = \Pr{U \in \calF}
\end{align*}
for every $G\in\calG$ and $\calF\subset\calG$. In other words, the Haar measure on $\calG$ is the normalized counting measure, i.e., $\mu_\calG(\{G\}) = \frac{1}{|\calG|}$ for every $G\in\calG$.

Let $\pi$ be a permutation of $\{1,\ldots,N\}$. The permutation matrix associated to $\pi$ is the $N\by N$ matrix given by $(\delta_{\pi(i),j})_{i,j=1}^N$. We denote by $\calS_N$ the set of all $N\by N$ permutation matrices. Also, we denote by $\Diag{N}{\pm}$ the set of $N\by N$ diagonal matrices with diagonal entries in $\{-1,1\}$. Observe that both $\calS_N$ and $\Diag{N}{\pm}$ are finite subgroups of $\Uni{N}$. Let $\calS_N^\pm$ be the set of $N\by N$ signed permutation matrices, i.e., $\calS_N^\pm = \calS_N \Diag{N}{\pm}$. If $P$ and $S$ are independent $N\by N$ random matrices distributed according to the Haar measure on $\calS_N$ and $\Diag{N}{\pm}$ respectively, then $PS$ is distributed according to the Haar measure on $\calS_N^\pm$. In this context we have the following lemma.\\

\noindent{\bf Lemma 4} For every $A\in\Mat{N}{\C}$, $\Av{\calS_N^\pm}{A}=\frac{\Tr{A}}{N}{\rm I}_N$. In particular,
\begin{equation*}
\Av{\calS_N^\pm}{\CvM{N,1}{\C}} = \left\{\frac{1}{N}{\rm I}_N\right\}.
\end{equation*}

\noindent{\it Proof.} See Appendix \ref{Appendix:Lemma4}. $\square$\\

If a channel has a group of symmetries that contains $\calS_N^\pm$, then the isotropic input is optimal for that channel (cf. Corollary 1). A non-trivial application of the previous lemma is Corollary 6 below.

\subsection{Block symmetries}
\label{Subsection:BlockSymmetries}

In the context of \cite{DiPe15}, a block symmetry is a matrix of the form ${\rm I}_d\otimes V$ with $V\in\Uni{N}$ such that $H({\rm I}_d\otimes V)\stackrel{\calL}{=}H$ where $H$ is the $dN\by dN$ matrix of propagation coefficients of the channel. In this paper we propose a broader scope for the notion of block symmetry: block symmetries are the tensor product or direct sum of elementary symmetries (unitary, diagonal or signed permutation). Even though block symmetries might seem artificial, they can be present in subtle ways; see Corollary 6 (cf. \cite[Th. 1]{VeSiVa03}).

Let $\calG_1$ and $\calG_2$ be closed subgroups of $\Uni{N_1}$ and $\Uni{N_2}$ respectively. Let $U_1$ and $U_2$ be independent random matrices distributed according to the Haar measure on $\calG_1$ and $\calG_2$ respectively. For $V_1\in\calG_1$ and $V_2\in\calG_2$,
\begin{equation*}
(U_1\otimes U_2)(V_1\otimes V_2) = U_1V_1 \otimes U_2V_2 \stackrel{\calL}{=} U_1\otimes U_2.
\end{equation*}
The multiplication invariance above shows that $U_1\otimes U_2$ is distributed according to the Haar measure on $\calG_1\otimes\calG_2\subset\calU(N_1N_2)$. Observe that $\calG_1\otimes\calG_2$ is closed, so the extended average operator with respect to this group is well defined.\\

\noindent{\bf Lemma 5} Let $\calG_1$ and $\calG_2$ be closed subgroups of $\Uni{N_1}$ and $\Uni{N_2}$ respectively. Let $\calG:=\calG_1\otimes\calG_2$ and $N:=N_1N_2$. If $\Av{\calG_2}{B} = \frac{\Tr{B}}{N_2}{\rm I}_{N_2}$ for all $B\in\Mat{N_2}{\C}$, then
\begin{equation*}
\Av{\calG}{\CvM{N,1}{\C}} = \Av{\calG_1}{\CvM{N_1,1}{\C}} \otimes \frac{1}{N_2}{\rm I}_{N_2}.
\end{equation*}

\noindent{\it Proof.} See Appendix \ref{Appendix:Lemma5}. $\square$\\

Lemma 2 and Lemma 5 allow us to easily recover one of the main results in \cite{DiPe15}.\\

\noindent{\bf Corollary 4} \cite[Th. 1]{DiPe15} Suppose that $H$ is a $dN \by dN$ random matrix such that $H({\rm I}_d \otimes W) \stackrel{\calL}{=} H$ for all $W\in\Uni{N}$. Then there is a CAICM for $H$ in the set $\CvM{d,1}{\C} \otimes \frac{1}{N}{\rm I}_N$.\\

\noindent{\it Proof.} The hypothesis clearly implies that ${\rm I}_d\otimes\Uni{N}\subset\calG(H)$. By Theorem 1, there is a CAICM for $H$ in
\begin{equation*}
\Av{\calG(H)}{\CvM{dN,1}{\C}} \subset \Av{{\rm I}_d\otimes\Uni{N}}{\CvM{N,1}{\C}}.
\end{equation*}
By Lemma 2,  $\Av{\Uni{N}}{B}=\frac{\Tr{B}}{N}{\rm I}_N$ for all $B\in\Mat{N}{\C}$. Thus, Lemma 5 implies that
\begin{equation*}
\Av{{\rm I}_d\otimes\Uni{N}}{\CvM{N,1}{\C}} = \Av{\{{\rm I}_d\}}{\CvM{d,1}{\C}} \otimes \frac{1}{N}{\rm I}_N.
\end{equation*}
Since the average operator with respect to $\{{\rm I}_d\}$ is the identity function on $\Mat{d}{\C}$, we conclude that there is a CAICM for $H$ in $\CvM{d,1}{\C} \otimes \frac{1}{N}{\rm I}_N$. $\square$\\

The machinery developed so far allows us to extend the previous corollary in a straightforward manner.\\

\noindent{\bf Corollary 5} Suppose that $H$ is a $dN\by dN$ random matrix such that $H(V\otimes W) \stackrel{\calL}{=} H$ for all $V\in\Diag{d}{\T}$ (resp. $V\in\Uni{d}$) and $W\in\Uni{N}$. Then there is a CAICM for $H$ in $\Diag{d,1}{\R_+}\otimes\frac{1}{N}{\rm I}_N$ (resp. $\{\frac{1}{dN}{\rm I}_{dN}\}$).\\

\noindent{\it Proof.} The proof follows the same steps as the proof of Corollary 4. The details are left to the reader. $\square$\\

Another natural family of block symmetries is obtained when instead of taking tensor products we take direct sums. For $1\leq k\leq K$, let $N_k$ be a positive integer and $\calG_k$ be a closed subgroup of $\Uni{N_k}$. Let $\calG=\bigoplus_{k=1}^K \calG_k\subset \Uni{N}$ where $N=\sum_{k=1}^K N_k$. Let $U_1,\ldots,U_K$ be independent random matrices such that $U_k$ is distributed according to the Haar measure on $\calG_k$ for every $1\leq k\leq K$. Let $U=\bigoplus_{k=1}^K U_k$. For $(G_1,\ldots,G_K)\in\calG_1\times\cdots\times\calG_K$,
\begin{equation*}
U\bigoplus_{k=1}^K G_k = \bigoplus_{k=1}^K U_kG_k \stackrel{\calL}{=} \bigoplus_{k=1}^K U_k = U.
\end{equation*}
The previous equality in distribution shows that $U$ is distributed according to the Haar measure on $\calG$. Observe that $\calG$ is closed, so the extended average operator with respect to this group is well defined.\\

\noindent{\bf Lemma 6} For $1\leq k\leq K$, let $N_k$ be a positive integer and $\calG_k$ be a closed subgroup of $\Uni{N_k}$. Let $\calG=\bigoplus_{k=1}^K \calG_k\subset \Uni{N}$ where $N=\sum_{k=1}^K N_k$. Assume that $\int_{\calG_k} F \dif\mu_{\calG_k}(F)$ is non-zero for at most one $k\in\{1,\ldots,K\}$. Then $\Av{\calG}{A}=\bigoplus_{k=1}^K \Av{\calG_k}{A^{(k,k)}}$ for every $A\in\Mat{N}{\C}$ where $A^{(i,j)}$ is the $N_i\by N_j$ matrix such that $A=(A^{(i,j)})_{i,j=1}^K$. In particular,
\begin{equation*}
\Av{\calG}{\CvM{N,1}{\C}} = \bigcup_{\begin{smallmatrix}p_1,\ldots,p_K\geq0\\p_1+\cdots+p_K=1\end{smallmatrix}} \bigoplus_{k=1}^K p_k\Av{\calG_k}{\CvM{N_k,1}{\C}}.
\end{equation*}

\noindent{\it Proof.} See Appendix \ref{Appendix:Lemma6}. $\square$\\

Lemmas 4 and 6 imply the following Corollary.\\

\noindent{\bf Corollary 6}  Let $H=\overline{H}+\tilde{H}$ where $\overline{H}$ is an $M\by N$ deterministic matrix and $\tilde{H}$ is an $M\by N$ Gaussian matrix. Let $\overline{H}=VEW^*$ be the SVD of $\overline{H}$. If
\begin{equation*}
\begin{matrix}
E_{1,1}=E_{2,2}=\cdots=E_{N_1,N_1},\\
E_{N_1+1,N_1+1}=E_{N_1+2,N_1+2}=\cdots=E_{N_1+N_2,N_1+N_2},\\
\vdots\\
E_{N_1+\cdots+N_{K-1}+1,N_1+\cdots+N_{K-1}+1}=\cdots=E_{N,N},
\end{matrix}
\end{equation*}
for some $N_1,\ldots,N_K\geq1$ with $N_1+\cdots+N_K=N$, then there is a CAICM for $H$ of the form $WDW^*$ for some $D\in\Diag{N,1}{\R_+}$ such that
\begin{equation*}
\begin{matrix}
D_{1,1}=D_{2,2}=\cdots=D_{N_1,N_1},\\
D_{N_1+1,N_1+1}=D_{N_1+2,N_1+2}=\cdots=D_{N_1+N_2,N_1+N_2},\\
\vdots\\
D_{N_1+\cdots+N_{K-1}+1,N_1+\cdots+N_{K-1}+1}=\cdots=D_{N,N}.
\end{matrix}
\end{equation*}

\noindent{\it Proof.} Observe that $H^*H = (EW^*+V^*\tilde{H})^* (EW^*+V^*\tilde{H})$. By the unitarily invariance of $\tilde{H}$ \cite[Lemma 5]{Te99}, we have that $H^*H\stackrel{\mathcal{L}}{=} (EW^*+\tilde{H})^* (EW^*+\tilde{H})$. In particular, this implies that $\calG(H) = \calG(EW^*+\tilde{H})$. Let $\calS=\bigoplus_{k=1}^K \calS_{N_k}^\pm$. By the assumption on $E$, we have that $S^*ES=E$ for all $S\in\calS$. Using this observation it can be shown that
\begin{align*}
(WSW^*)^*(EW^*+\tilde{H})^*(EW^*+\tilde{H})(WSW^*) &= (EW^*+S^*\tilde{H}WSW^*)^*(EW^*+S^*\tilde{H}WSW^*)\\
&\stackrel{\mathcal{L}}{=}(EW^*+\tilde{H})^*(EW^*+\tilde{H})
\end{align*}
for all $S\in\calS$. This shows that $W\calS W^*\subset \calG(H)$ and so, by Theorem 1, there is a CAICM for $H$ in $\Av{W\calS W^*}{\CvM{N,1}{\C}}$. It is easy to show that, for $A\in\Mat{N}{\C}$,
\begin{align*}
\Av{W\calS W^*}{A} = W\Av{\calS}{W^*AW}W^*.
\end{align*}
In particular, $\Av{W\calS W^*}{\CvM{N,1}{\C}}=W\Av{\calS}{\CvM{N,1}{\C}}W^*$. By Lemma 6, we have then
\begin{equation*}
\Av{W\calS W^*}{\CvM{N,1}{\C}} = W \left(\bigcup_{\begin{smallmatrix}p_1,\ldots,p_K\geq0\\p_1+\cdots+p_K=1\end{smallmatrix}}\bigoplus_{k=1}^K p_k\Av{\calS_{N_k}^\pm}{\CvM{N_k,1}{\C}}\right) W^*.
\end{equation*}
By Lemma 4, we conclude that
\begin{align*}
\Av{W\calS W^*}{\CvM{N,1}{\C}} &= \left\{W\left(\bigoplus_{k=1}^K \frac{p_k}{N_k}{\rm I}_{N_k}\right) W^* \bigg| \begin{matrix}p_1,\ldots,p_K\geq0,\\p_1+\cdots+p_K=1\end{matrix}\right\}.
\end{align*}
The result follows from the previous equation. $\square$\\

The previous corollary recovers the fact that the isotropic input is optimal for the standard Ricean channel when $E_{1,1}=\cdots=E_{N,N}$ \cite{LoTuVe06}. The main theorem in \cite{VeSiVa03} can be recover from the previous corollary setting $N_k=1$ for all $1\leq k < K$ and $N_K\geq 1$.

\section{Optimality of the Main Theorem}
\label{Section:OptimalityMainTheorem}

In this section, we exhibit a set of channels such that: a) they have the same group of symmetries $\calG$, b) each one has a unique CAICM, and c) for every $Q\in\Av{\calG}{\CvM{N,1}{\C}}$ there is a channel with $Q$ as its CAICM. This will show that in general, based solely on the group of symmetries $\calG(H)$ of a channel $H$, we cannot say more about a CAICM for $H$ than being in the set $\Av{\calG(H)}{\CvM{N,1}{\C}}$, as stablished in Theorem 1.

For $\alpha\geq\sqrt{2}^{-1}$, let
\begin{equation*}
H_\alpha := \left(\begin{matrix}1&0\\0&\alpha v\end{matrix}\right)
\end{equation*}
where $v$ is a random variable uniformly distributed on $\T$. A direct computation shows that
\begin{equation*}
\calG(H_\alpha) = \left\{\left(\begin{matrix}1&0\\0&t\end{matrix}\right) \bigg| t\in\T\right\} =: \calG.
\end{equation*}
For $\left(\begin{smallmatrix}a&\overline{c}\\c&b\end{smallmatrix}\right)\in\CvM{2,1}{\C}$, we have that $I_{H_\alpha}\left(\left(\begin{smallmatrix}a&\overline{c}\\c&b\end{smallmatrix}\right)\right) = \log\left[(1+a)(1+\alpha^2b)-\alpha^2|c|^2\right]$. A standard optimization argument shows that $C_{H_\alpha} = 2\log\left(\frac{1+2\alpha^2}{2\alpha}\right)$ and that there is a unique CAICM for $H_\alpha$ determined by $\hat{a} = \frac{1}{2\alpha^2}$ and $\hat{c}=0$.

Similarly, let
\begin{equation*}
H_\infty=\left(\begin{matrix}0&0\\1&2v\end{matrix}\right).
\end{equation*}
In this case, the group of symmetries of $H_\infty$ is also $\calG$. A direct computation shows that, for $\left(\begin{smallmatrix}a&\overline{c}\\c&b\end{smallmatrix}\right)\in\CvM{2,1}{\C}$, $I_{H_\infty}\left(\left(\begin{smallmatrix}a&\overline{c}\\c&b\end{smallmatrix}\right)\right) = \E{\log(1+a+4b+4\Re(cv))}$. After an application of Jensen's inequality, we obtain that $C_{H_\infty}=\log(5)$ and that there is a unique CAICM for $H_\infty$ determined by $\hat{a}=0$ and $\hat{c}=0$.

By a multiplication invariance argument, it can be shown that $\left(\begin{smallmatrix}1&0\\0&v\end{smallmatrix}\right)$ is distributed according to the Haar measure on $\calG$. Thus, a direct computation shows that
\begin{equation*}
\Av{\calG}{\CvM{2,1}{\C}} = \left\{\left(\begin{matrix}a&0\\0&b\end{matrix}\right) \bigg| a,b\geq0;a+b=1\right\}.
\end{equation*}
From this it is clear that the family $\{H_\alpha \mid \sqrt{2}^{-1}\leq \alpha\leq \infty\}$ has the desired properties described at the beginning of this section.

\section{Conclusion}
\label{Section:Conclusion}

In this paper we found an approach to study the capacity achieving input covariance matrices of single user multiantenna channels based solely on the group of symmetries of their matrices of propagation coefficients. This allowed us to unify and improve the techniques used in many classical theorems. Our main result provided us a set where a capacity achieving input covariance matrix is guaranteed to exist. Since this set was obtained using the Haar (uniform) measure on the group of symmetries, it avoided any ingenious guessing to apply it, as opposed to the classical approach, and produced a set smaller than or equal to those obtained by any ingenious guessing. Our main theorem led to simple proofs for old and new capacity theorems. Among other results, we showed that in a channel with at least two standard symmetries, the isotropic input is optimal. Overall, we made explicit the fundamental connection between the capacity achieving input covariance matrices of a channel and the group of symmetries of its matrix of propagation coefficients.

\appendix

\section{Proof of Lemma 1}
\label{Appendix:Lemma1}

By the definition of the group of symmetries (\ref{eq:DefGroupSymmetries}), it is clear that $\calG(H)\subset\Uni{N}$. The equality ${\rm I}_N^* (H^*H) {\rm I}_N = H^*H$ implies that ${\rm I}_N\in\calG(H)$, and thus $\calG(H)$ is not empty. If $G_1,G_2\in\calG(H)$, then
\begin{align*}
H^*H &\stackrel{\calL}{=} G_2^* (H^*H) G_2 \stackrel{\calL}{=} G_2^*(G_1^*(H^*H)G_1)G_2 = (G_1G_2)^* (H^*H) (G_1G_2),
\end{align*}
i.e., $G_1G_2\in\calG(H)$. Similarly, if $G\in\calG(H)$, then
\begin{align*}
H^*H &= (GG^{-1})^*(H^*H)GG^{-1} \stackrel{\calL}{=} G^{-1\ast} (H^*H) G^{-1},
\end{align*}
i.e., $G^{-1}\in\calG(H)$. Therefore $\calG(H)$ is a non-empty subgroup of $\Uni{N}$. It remains to show that $\calG(H)$ is closed.

Let $G\in\Uni{N}$ be a limit point of $\calG(H)$. Thus, there exists $\{G_n\}_{n\geq1}\subset\calG(H)$ such that $G_n\to G$. Let $\calO$ be an open subset of $\Mat{M\by N}{\C}$. By Fatou's lemma and the fact that $\{G_n\}_{n\geq1}\subset\calG(H)$,
\begin{align*}
\E{\liminf_{n\to\infty} \I{G_n^*(H^*H) G_n\in\calO}} &\leq \liminf_{n\to\infty} \E{\I{G_n^*(H^*H)G_n\in\calO}}\\
&= \E{\I{H^*H\in\calO}}.
\end{align*}
Suppose that $G^*AG\in\calO$ for some $A\in\Mat{M\by N}{\C}$. Since $G_n\to G$, $\displaystyle \lim_{n\to\infty} G_n^*AG_n=G^*AG$. Since $\calO$ is open, the latter implies that $G_n^*AG_n\in\calO$ for $n$ big enough. In particular, $\I{G^*(H^*H)G\in\calO}\leq\liminf_{n\to\infty} \I{G_n^*(H^*H)G_n\in\calO}$ and therefore $\E{\I{G^*(H^*H)G\in\calO}} \leq \E{\I{H^*H\in\calO}}$, i.e.,
\begin{equation}
\label{eq:ProofLemma1}
\Pr{G^*(H^*H)G\in\calO} \leq \Pr{H^*H\in\calO}.
\end{equation}
For $X\in\Mat{M\by N}{\C}$ and $\rho>0$, let $B(X,\rho) := \{A\in\Mat{M\by N}{\C} \mid \|X-A\|<\rho\}$. Consider $\calB=B(C,r)$ for some $C\in\Mat{M\by N}{\C}$ and $r>0$. By the continuity of the probability \cite[Th. 1, Ch. 1]{HoPoSt71}, equation (\ref{eq:ProofLemma1}) implies that
\begin{align}
\nonumber \Pr{G(H^*H)G\in\overline{\calB}} &= \lim_{k\to\infty} \Pr{G^*(H^*H)G\in B(C,r+k^{-1})}\\
\nonumber &\leq \lim_{k\to\infty} \Pr{H^*H\in B(C,r+k^{-1})}\\
\label{eq:Lemma1ProbClosed} &= \Pr{H^*H\in\overline{\calB}}.
\end{align}
Applying equation (\ref{eq:ProofLemma1}) to $\overline{\calB}^c$, we obtain that
\begin{equation}
\label{eq:Lemma1ProbOpen}
\Pr{G^*(H^*H)G\in\overline{\calB}^c} \leq \Pr{H^*H\in\overline{\calB}^c}.
\end{equation}
Since
\begin{align*}
1 &= \Pr{G^*(H^*H)G\in\overline{\calB}}+\Pr{G^*(H^*H)G\in\overline{\calB}^c}\\
&= \Pr{H^*H\in\overline{\calB}}+\Pr{H^*H\in\overline{\calB}^c},
\end{align*}
the inequalities (\ref{eq:Lemma1ProbClosed}) and (\ref{eq:Lemma1ProbOpen}) imply that $\Pr{HG\in\overline{\calB}} = \Pr{H\in\overline{\calB}}$. The previous equality can be extended to any Borel set by standard arguments, showing then that $G\in\calG(H)$ and therefore $\calG(H)$ is closed.

\section{Proof of Proposition 4}
\label{Appendix:Prop4}

a) $\Rightarrow$ b). Suppose that $\{Q_n\}_{n\geq1}\subset\CvM{N,1}{\C}$ converge to $Q\in\CvM{N,1}{\C}$. Let $Z_n:=\log\det({\rm I}_M+HQ_nH^*)$ for $n\geq1$ and $Z:=\log\det({\rm I}_M+HQH^*)$. Since $Q_n\to Q$, we have that $Z_n\to Z$ pointwise and, in particular, almost surely. If we can show that there exists $Y$ such that $0\leq Z_n\leq Y$ for all $n\geq1$ and $\E{Y}<\infty$, then by the dominated convergence theorem we will conclude that $\lim_{n\to\infty} I_H(Q_n) = I_H(Q)$. In particular, this will show that $I_H(\cdot)$ is continuous on $\CvM{N,1}{\C}$. Let us prove that such a $Y$ in fact exists.

Let $n\geq1$, then
\begin{align*}
Z_n = M \frac{1}{M} \sum_{k=1}^M \log(1+\lambda_k(HQ_nH^*))
\end{align*}
where $\lambda_1(HQ_nH^*)\geq\cdots\geq\lambda_M(HQ_nH^*)\geq0$ are the eigenvalues of $HQ_nH^*$. By Jensen's inequality,
\begin{align*}
Z_n &\leq M \log\left(1+ \frac{1}{M} \sum_{k=1}^M \lambda_k(HQ_nH^*)\right) = M\log\left(1+\frac{\Tr{HQH^*}}{M}\right).
\end{align*}
Since $Q\in\CvM{N,1}{\C}$, $|Q_{i,j}|\leq1$ for all $1\leq i,j\leq N$. Using the arithmetic mean-quadratic mean inequality, we obtain that
\begin{align*}
\Tr{HQH^*} &= \sum_{i=1}^M \sum_{j,k=1}^N H_{i,j} Q_{j,k} \overline{H_{i,k}} \leq \sum_{i=1}^M \sum_{j,k=1}^N |H_{i,j}|\ |H_{i,k}|\\
&= \sum_{i=1}^M \left(\sum_{j=1}^N |H_{i,j}|\right)^2 \leq N \sum_{i=1}^M \sum_{j=1}^N |H_{i,j}|^2=N \|H\|^2.
\end{align*}
By the monotonicity of the logarithm,
\begin{equation*}
Z_n \leq M \log\left(1+\frac{N}{M}\|H\|^2\right) \leq 2M \log\left(1+\sqrt{\frac{N}{M}}\|H\|\right).
\end{equation*}
Recall that $\log(1+\alpha x) \leq \log(1+x) + \log(1+\alpha)$ for $\alpha,x\geq0$. Thus,
\begin{equation*}
Z_n \leq 2M\log(1+\|H\|)+2M\log\left(1+\sqrt{\frac{N}{M}}\right)=:Y.
\end{equation*}
The hypothesis $\E{\log(1+\|H\|)}<\infty$ immediately implies that $\E{Y}<\infty$.

b) $\Rightarrow$ c). The fact that $C_H$ is finite follows immediately from the continuity of $I_H(\cdot)$ and the compactness of $\CvM{N,1}{\C}$.

c) $\Rightarrow$ a). Using the fact that $1+\sqrt{x}\leq2(1+x)$ for all $x\geq0$ and the monotonicity of the logarithm,
\begin{align*}
\E{\log(1+\|H\|)} &= \E{\log\left(1+\sqrt{\Tr{H^*H}}\right)}\\
&\leq \E{\log\left(1+N\Tr{\frac{HH^*}{N}}\right)} +\log(2)\\
&\leq \E{\log\left(1+\Tr{\frac{HH^*}{N}}\right)} + \log(2N)\\
&= \E{\log\left(1+\sum_{k=1}^M \lambda_k(HH^*/N)\right)}+\log(2N),
\end{align*}
where $\lambda_1(HH^*/N)\geq\cdots\geq\lambda_M(HH^*/N)\geq0$ are the eigenvalues of $HH^*/N$. Using the inequality $1+x_1+\cdots+x_M\leq (1+x_1)\cdots(1+x_M)$ for $x_1,\ldots,x_M\geq0$,
\begin{align}
\label{eq:InqBoundMutualInformation} \E{\log(1+\|H\|)} & \leq \E{\log\det\left({\rm I}_M+\frac{HH^*}{N}\right)}+\log(2N)\\
\nonumber &= I_H(N^{-1}{\rm I}_N) + \log(2N) \leq C_H + \log(2N).
\end{align}
Since $C_H$ is finite, we conclude that $\E{\log(1+\|H\|)}$ is also finite.

If $C_H=\infty$, by the equivalence between a) and c),  then $\E{\log(1+\|H\|)}=\infty$. By the inequality in (\ref{eq:InqBoundMutualInformation}), we conclude that $I_H\left(\frac{1}{N}{\rm I}_N\right)=\infty$. The converse is clear.

\section{Proof of Theorem 1}
\label{Appendix:Thm1}

a) If $C_H=\infty$, by Proposition 4 we know that the normalized identity achieves capacity. Since $\frac{1}{N}{\rm I}_N=\Av{\calG(H)}{\frac{1}{N}{\rm I}_N}\in\Av{\calG(H)}{\CvM{N,1}{\C}}$, the result follows in this case.

Assume otherwise that $C_H$ is finite. Let $Q\in\CvM{N,1}{\C}$. Then
\begin{align*}
I_H(\Av{\calG(H)}{Q}) = \E{\log\det\left({\rm I}_M + H \int_{\calG(H)} GQG^* \dif\mu_{\calG(H)}(G) H^*\right)}.
\end{align*}
Let $\mu_*$ be the pushforward measure of $\mu_{\calG(H)}$ by $G\mapsto GQG^*$. Then, by the change of variable formula,
\begin{align*}
I_H(\Av{\calG(H)}{Q}) = \E{\log\det\left({\rm I}_M+H\int_{\CvM{N,1}{\C}} \Phi \dif\mu_*(\Phi) H^*\right)}.
\end{align*}
Since the mapping $\Phi \mapsto \log\det({\rm I}_M + H\Phi H^*)$ is concave \cite{Te99}, Jensen's inequality implies that
\begin{align*}
&I_H(\Av{\calG(H)}{Q}) \geq \E{\int_{\CvM{N}{\C}{1}} \log\det\left({\rm I}_M+H\Phi H^*\right) \dif\mu_*(\Phi)}.
\end{align*}
By the change of variable formula and Tonelli's theorem,
\begin{align*}
I_H(\Av{\calG(H)}{Q}) & \geq \E{\int_{\calG(H)} \log\det\left({\rm I}_M+HGQG^* H^*\right) \dif\mu_{\calG(H)}(G)}\\
&= \int_{\calG(H)} \E{\log\det\left({\rm I}_M+(HG)QG^*H^*\right)} \dif\mu_{\calG(H)}(G).
\end{align*}
By the identity $\det({\rm I}+AB) = \det({\rm I}+BA)$ \cite[Sec. 4.33]{Se2008}, we have that
\begin{equation*}
I_H(\Av{\calG(H)}{Q}) \geq \int_{\calG(H)} \E{\log\det\left({\rm I}_N+QG^*(H^*H)G\right)} \dif\mu_{\calG(H)}(G).
\end{equation*}
Since $G^*(H^*H)G\stackrel{\calL}{=}H^*H$ for all $G\in\calG(H)$, the same identity implies that
\begin{align*}
I_H(\Av{\calG(H)}{Q}) &\geq \int_{\calG(H)} \E{\log\det\left(I_M+HQH^*\right)} \dif\mu_{\calG(H)}(G) =  I_H(Q).
\end{align*}
Since we are assuming finite capacity, Proposition 4 implies that $I_H(\cdot)$ is continuous on $\CvM{N,1}{\C}$. By compactness of $\CvM{N,1}{\C}$ we conclude that there exists a CAICM $Q_0\in\CvM{N,1}{\C}$. By the previous inequality we have then that $\Av{\calG(H)}{Q_0}$ is also a CAICM for $H$. This establishes that there exists a CAICM for $H$ in $\Av{\calG(H)}{\CvM{N,1}{\C}}$.

b) Assume that $\calF$ is a closed subgroup of $\calG(H)$. If $F\in\calG(H)$ then, by the multiplication invariance property of $\mu_{\calG(H)}$,
\begin{equation*}
F \Av{\calG(H)}{Q'}F^* = \E{FUQ'(FU)^*} = \Av{\calG(H)}{Q'},
\end{equation*}
where $Q'\in\CvM{N,1}{\C}$ and $U$ is a random matrix distributed according to the Haar measure on $\calG(H)$. If $Q=\Av{\calG(H)}{Q'}\in\Av{\calG(H)}{\CvM{N,1}{\C}}$ then
\begin{align*}
\Av{\calF}{Q} &= \int_\calF F\Av{\calG(H)}{Q'}F^* \dif\mu_\calH(F) = \Av{\calG(H)}{Q'} = Q.
\end{align*}
This implies that $Q\in\Av{\calF}{\CvM{N,1}{\C}}$ and thus $\Av{\calG(H)}{\CvM{N,1}{\C}} \subset \Av{\calF}{\CvM{N,1}{\C}}$, as required. The proof when $\calF$ is a finite multiset with elements in $\calG(H)$ is proved analogously. The details are left to the reader.

\section{Proof of Proposition 2}
\label{Appendix:Prop2}

By Theorem 1 part b), $\Av{\calG(H)}{\CvM{N,1}{\C}}\subset\Av{\calF_k}{\CvM{N,1}{\C}}$ for all $1\leq k\leq K$. Thus, $\Av{\calG(H)}{\CvM{N,1}{\C}}\subset\bigcap_{k=1}^K \Av{\calF_k}{\CvM{N,1}{\C}}$. By Theorem 1 part a) there is a CAICM for $H$ in $\Av{\calG(H)}{\CvM{N,1}{\C}}$ and therefore in $\bigcap_{k=1}^K \Av{\calF_k}{\CvM{N,1}{\C}}$.

\section{Proof of Lemma 2}
\label{Appendix:Lemma2}

The fact that $\Av{\Uni{N}}{A}=\frac{\Tr{A}}{N}{\rm I}_N$ is a direct application of equations (\ref{eq:AvMoments}) and (\ref{eq:MomentsHaarUnitary}). This implies that $\Av{\Uni{N}}{Q}=\frac{1}{N}{\rm I}_N$ for every $Q\in\CvM{N,1}{\C}$, i.e., $\Av{\Uni{N}}{\CvM{N,1}{\C}}=\{\frac{1}{N}{\rm I}_N\}$.

\section{Proof of Lemma 3}
\label{Appendix:Lemma3}

Recall the description of the Haar measure on $W\Diag{N}{\T}W^*$ obtained at the beginning of Section \ref{Subsection:DiagonalSymmetries}. In particular, for every $A\in\Mat{N}{\C}$,
\begin{align*}
\Av{W\Diag{N}{\T}W^*}{A} &= \E{WUW^*AWU^*W^*} = W\E{U(W^*AW)U^*}W^*.
\end{align*}
By the independence of $u_1,\ldots,u_N$ and the fact that $\E{u_n}=0$ for all $1\leq n\leq N$, it is straightforward to verify that $\E{U(W^*AW)U^*}=\Delta(W^*AW)$. Therefore,
\begin{equation*}
\Av{W\Diag{N}{\T}W^*}{A} = W\Delta(W^*AW)W^*.
\end{equation*}

It is easy to verify from the previous equation that $\Av{W\Diag{N}{\T}W^*}{\CvM{N,1}{\C}} \subset W\Diag{N,1}{\R_+}W^*$. If $WDW^*\in W\Diag{N,1}{\R_+}W^*$, then
\begin{equation*}
\Av{W\Diag{N}{\T}W^*}{WDW^*} = W\Delta(D)W^* = WDW^*.
\end{equation*}
Since $W\Diag{N,1}{\R_+}W^* \subset \CvM{N,1}{\C}$, the latter equation implies that $WDW^*\in \Av{W\Diag{N}{\T}W^*}{\CvM{N,1}{\C}}$. This proves the inclusion $W\Diag{N,1}{\R_+}W^* \subset \Av{W\Diag{N}{\T}W^*}{\CvM{N,1}{\C}}$, and thus the required equality.

\section{Proof of Lemma 4}
\label{Appendix:Lemma4}

Recall the description of the Haar measure on $\calS_N^\pm$ given in Section \ref{Subsection:SignedPermutationSymmetries}. In particular, $\Av{\calS_N^\pm}{A} = \E{PSASP^*}$ for every $A\in\Mat{N}{\C}$. Since $P$ and $S$ are independent, we obtain that
\begin{align*}
\Av{\calS_N^\pm}{A} &= \E{P\E{SAS|P}P^*} = \E{P\E{SAS}P^T} = \Av{\calS_N}{\Av{\Diag{N}{\pm}}{A}}.
\end{align*}
Similarly to the proof of Lemma 3, it can be shown that $\Av{\Diag{N}{\pm}}{A}=\Delta(A)$. By \cite[Lemma 1]{DiPe15}, for every $B\in\Mat{N}{\C}$,
\begin{equation*}
\Av{\calS_N}{B} = \frac{\Tr{B}}{N}{\rm I}_N + \left(\frac{1}{N(N-1)}\sum_{i\neq j} B_{i,j}\right) ({\rm J}_N - {\rm I}_N)
\end{equation*}
where $J_N$ is the $N\by N$ matrix with all its entries equal to one. Therefore, we conclude that
\begin{align*}
\Av{\calS_N^\pm}{A} &= \Av{\calS_N}{\Delta(A)} = \frac{\Tr{\Delta(A)}}{N}{\rm I}_N = \frac{\Tr{A}}{N}{\rm I}_N.
\end{align*}
By the previous equation, the rest of the proof follows the same steps as in Lemma 2.

\section{Proof of Lemma 5}
\label{Appendix:Lemma5}

Let $Q\in\Av{\calG}{\CvM{N,1}{\C}}$, i.e., $Q=\Av{\calG}{Q'}$ for some $Q'\in\CvM{N,1}{\C}$. Then there exists $K\geq1$ such that $Q'=\sum_{k=1}^K A_k\otimes B_k$ for some $A_1,\ldots,A_K\in\Mat{N_1}{\C}$ and $B_1,\ldots,B_K\in\Mat{N_2}{\C}$. Recall the description for the Haar measure on $\calG_1\otimes\calG_2$ given in Section \ref{Subsection:BlockSymmetries}. In particular, we have that
\begin{align*}
\Av{\calG}{Q'} &= \E{(U_1\otimes U_2) Q' (U_1^*\otimes U_2^*)} = \sum_{k=1}^K \E{U_1A_kU_1^*\otimes U_2B_kU_2^*}.
\end{align*}
By independence of $U_1$ and $U_2$, and the linearity of the average operator,
\begin{align*}
Q 
&= \sum_{k=1}^K \Av{\calG_1}{A_k}\otimes\Av{\calG_2}{B_k}\\
&= \sum_{k=1}^K \Av{\calG_1}{A_k} \otimes \frac{\Tr{B_k}}{N_2}{\rm I}_{N_2}\\
&= \Av{\calG_1}{\sum_{k=1}^K \Tr{B_k}A_k}\otimes\frac{1}{N_2}{\rm I}_{N_2}.
\end{align*}
Since $Q=\Av{\calG}{Q'}\in\CvM{N,1}{\C}$, the previous equation implies that $A\in\CvM{N_1,1}{\C}$ where
\begin{equation*}
A:=\Av{\calG_1}{\sum_{k=1}^K \Tr{B_k}A_k}.
\end{equation*}
A direct computation shows that $\AV_{\calG_1} = \AV_{\calG_1}\circ\AV_{\calG_1}$. Thus $Q  = \Av{\calG_1}{A}\otimes\frac{1}{N_2}{\rm I}_{N_2}$.
This proves that $\Av{\calG}{\CvM{N,1}{\C}}\subset\Av{\calG_1}{\CvM{N_1,1}{\C}} \otimes \frac{1}{N_2}{\rm I}_{N_2}$.

Conversely, let $Q\in\Av{\calG_1}{\CvM{N_1,1}{\C}}\otimes\frac{1}{N_2}{\rm I}_{N_2}$, i.e., $Q=\Av{\calG_1}{Q'}\otimes\frac{1}{N_2}{\rm I}_{N_2}$ for some $Q'\in\CvM{N_1,1}{\C}$. Then, $Q = \Av{\calG_1}{Q'}\otimes\frac{1}{N_2}{\rm I}_{N_2} = \Av{\calG}{Q'\otimes\frac{1}{N_2}{\rm I}_{N_2}}$. Since $Q'\otimes\frac{1}{N_2}{\rm I}_{N_2}\in\CvM{N,1}{\C}$, the previous equation proves that $\Av{\calG_1}{\CvM{N_1,1}{\C}}\otimes\frac{1}{N_2}{\rm I}_{N_2}\subset\Av{\calG}{\CvM{N,1}{\C}}$.

\section{Proof of Lemma 6}
\label{Appendix:Lemma6}

Recall the description for the Haar measure on $\calG=\bigoplus_{k=1}^K \calG_k$ given in Section \ref{Subsection:BlockSymmetries}. In this case,
\begin{align*}
\Av{\calG}{A} &= \E{UAU^*} = \left(\E{U_iA^{(i,j)}U_j}\right)_{i,j=1}^K.
\end{align*}
By assumption, $\E{U_k}$ is non-vanishing for at most one $k\in\{1,\ldots,K\}$. By the independence of $U_1,\ldots,U_K$, we conclude that
\begin{align*}
\Av{\calG}{A} &= \left(\delta_{i,j}\E{U_iA^{(i,j)}U_j}\right)_{i,j=1}^K = \left(\delta_{i,j}\Av{\calG_i}{A^{(i,i)}}\right)_{i,j=1}^K = \bigoplus_{k=1}^K \Av{\calG_k}{A^{(k,k)}}.
\end{align*}
For a covariance matrix $A\in\CvM{N,1}{\C}$, its diagonal submatrices $A^{(1,1)},\ldots,A^{(K,K)}$ are also covariance matrices with trace at most one. From this the inclusion
\begin{equation*}
\Av{\calG}{\CvM{N,1}{\C}} \subset \bigcup_{\begin{smallmatrix}p_1,\ldots,p_K\geq0\\p_1+\cdots+p_K=1\end{smallmatrix}}\bigoplus_{k=1}^K p_k\Av{\calG_k}{\CvM{N_k,1}{\C}}
\end{equation*}
is clear. The proof of the reversed inclusion follows the same steps as the last parts of Lemma 3 and 5.

\section{Proof of Proposition 3}
\label{Appendix:Prop3}

Suppose that $V$ is a standard symmetry and denote by $\overline{\langle V\rangle}$ the closure of the (multiplicative) group generated by $V$. It is known that $\overline{\langle V\rangle} = W\Diag{N}{\T}W^*$ \cite[Prop. 1.4.1]{KaHa95}. Let $\calF_1=\overline{\langle V_1\rangle}$ and $\calF_2=\overline{\langle V_2\rangle}$. By Lemma 3, for $k=1,2$, $\Av{\calF_k}{\CvM{N,1}{\C}} = W_k \Diag{N,1}{\R_+} W_k^*$. If we can show that $W_1 \Diag{N,1}{\R_+} W_1^* \cap W_2 \Diag{N,1}{\R_+} W_2^* = \left\{\frac{1}{N}{\rm I}_N\right\}$, then Proposition 2 will imply that the normalized identity achieves capacity.

Let $M\in W_1 \Diag{N,1}{\R_+} W_1^* \cap W_2 \Diag{N,1}{\R_+} W_2^*$. In particular, there exist $E_1,E_2\in\Diag{N,1}{\R_+}$ such that $M=W_1E_1W_1^*=W_2E_2W_2^*$. Equivalently, $E_1W=WE_2$. This implies that, for $1\leq i,j\leq N$, $(E_1)_{i,i}W_{i,j} = W_{i,j}(E_2)_{j,j}$. Since $W_{i,j}\neq0$, we conclude that $(E_1)_{i,i}=(E_2)_{j,j}$. The latter equality implies that $E_1=E_2=\frac{1}{N}{\rm I}_N$ and therefore $M=\frac{1}{N}{\rm I}_N$.

\section*{Acknowledgment}

I would like to thank J. Mingo and S. Asoodeh for the useful discussions held while preparing this paper. The comments and suggestions made by the reviewers are also acknowledged.

\end{document}